# High Photovoltaic Quantum Efficiency in Ultrathin van der Waals Heterostructures

*Joeson Wong[1‡], Deep Jariwala[1,2‡], Giulia Tagliabue[1,4], Kevin Tat[1], Artur R. Davoyan[1,2,3], Michelle C. Sherrott[1,2] and Harry A. Atwater[1,2,3,4*]*

[1]Department of Applied Physics and Materials Science, California Institute of Technology, Pasadena, CA-91125, USA

[2]Resnick Sustainability Institute, California Institute of Technology, Pasadena, CA-91125, USA

[3]Kavli Nanoscience Institute, California Institute of Technology, Pasadena, CA-91125, USA

[4]Joint Center for Artificial Photosynthesis, California Institute of Technology, Pasadena, CA-91125, USA

* Corresponding author: Harry A Atwater (haa@caltech.edu)

‡ These authors contributed equally

## ABSTRACT:

We report experimental measurements for ultrathin (< 15 nm) van der Waals heterostructures exhibiting external quantum efficiencies exceeding 50%, and show that these structures can achieve experimental absorbance > 90%. By coupling electromagnetic simulations and experimental measurements, we show that pn $WSe_2/MoS_2$ heterojunctions with vertical carrier collection can have internal photocarrier collection efficiencies exceeding 70%.

## GRAPHICAL TABLE OF CONTENTS:

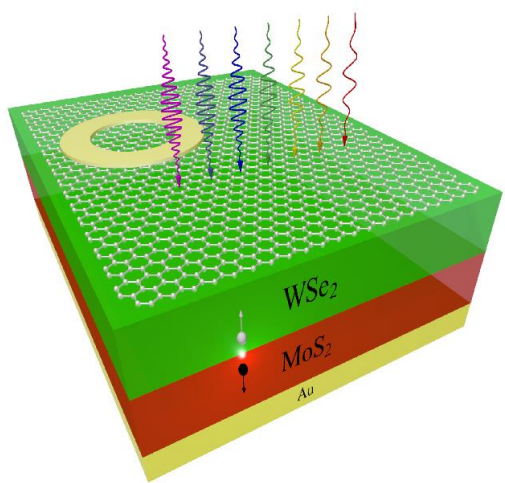
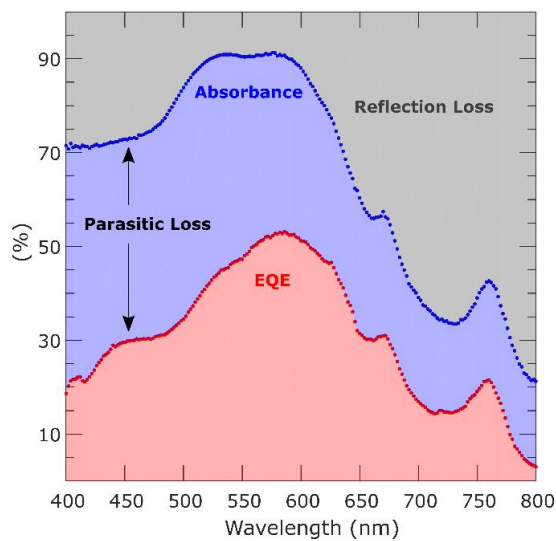

Owing to their naturally passivated basal planes and strong light-matter interactions, transition metal dichalcogenides are of considerable interest as active elements of optoelectronic devices such as light-emitting devices, photodetectors and photovoltaics.[1,2] Ultrathin transition metal dichalcogenide (TMD) photovoltaic devices a few atomic layers in thickness have been realized using TMDs such as molybdenum disulfide ($MoS_2$) and tungsten diselenide ($WSe_2$).[3–8] Complete absorption of the solar spectrum is a challenge as the thickness is reduced to the ultrathin limit,[9–11] whereas efficient carrier collection is challenging in thicker bulk TMD crystals. The active layers in conventional photovoltaics typically range from a few microns in direct gap materials (gallium arsenide) to a hundred microns thick or more in indirect gap materials (silicon).[12]

Efficient ultrathin and ultralight (<100 g/m$^2$) photovoltaics have long been sought for many applications where weight and flexibility are important design considerations, such as applications in space power systems, internet-of-things devices, as well as portable and flexible electronics.[13–15] Conventional photovoltaic materials are mechanically fragile when thinned down to the ultrathin (< 10 nm) regime, and interfacial reactions mean that a large fraction of the crystal consists of surface-modified regions rather than intrinsically bulk material. Surface oxides and dangling bonds in ultrathin films often result in increased nonradiative recombination losses, lowering photovoltaic efficiencies. By contrast, transition metal dichalcogenides have intrinsically high absorption and their layered crystallographic structures suggest the possibility of achieving intrinsically passive basal planes in high quality crystals.

Photovoltaics that can approach the Shockley-Queisser limit,[16,17] have two prerequisites: first, that at open circuit, every above-bandgap photon that is absorbed is extracted as an emitted photon at the band-edge of the material, *i.e.* it has perfect external radiative efficiency.[18] Amani *et al.* have recently demonstrated that superacid-treated monolayers of $MoS_2$ and $WS_2$ exhibit internal radiative efficiency > 99%,[19] suggesting that the condition of very high external radiative efficiency might be satisfied in transition metal dichalcogenides. The second prerequisite is that at short circuit, the photovoltaic device must convert every incident above-bandgap photon into an extracted electron, *i.e.* it has external quantum efficiency (EQE) approaching unity.

To understand the path to high EQE, we can deconvolute the external quantum efficiency into the product of two terms: the absorbance and internal quantum efficiency (IQE). High EQE devices exhibit both high absorption and internal quantum efficiency, *i.e.*, carrier generation and collection efficiency per absorbed photon. To date, reports of van der Waals based photovoltaic devices have not considered both of these concepts and separately evaluated them as criteria for high efficiency photovoltaics.

Coupling electromagnetic simulations with absorption and EQE measurements enables quantitative characterization of few-atomic-layer thickness optoelectronic devices in van der Waals heterostructures. In this paper, we demonstrate external quantum efficiencies > 50% (Figure 1(a)), indicating that van der Waals heterostructures have considerable potential for efficient photovoltaics. We show that high EQE results from both high optical absorption and efficient electronic charge carrier collection. We analyze the optical response using electromagnetic simulations to explain how near-unity absorption can be achieved in heterostructures (Figure 1(b)). We find that experimental absorption results for van der Waals heterostructures match well with these electromagnetic simulations. Thus we can separate optical absorption and electronic transport to quantitatively compare their effects on charge collection efficiency for both pn heterojunctions and Schottky junctions (Figure 1 (c)). In addition, we analyze the role of few-layer graphene as a transparent top contact (Figure 1 (d)). Finally, we outline important considerations for designing high efficiency photovoltaic devices. By simultaneously maximizing both external radiative efficiency and external quantum efficiency in a single device, van der Waals materials based photovoltaic devices could in principle achieve efficiencies close to the Shockley-Queisser limit for their bandgaps.

## RESULTS AND DISCUSSION

## Optoelectronic device characteristics

We analyzed the optoelectronic device characteristics of a high-performance device consisting of a vertical van der Waals heterostructure device of 0.6 nm thick few-layer graphene (FLG)/9 nm $WSe_2$/3 nm $MoS_2$/Au (see Supporting Information S1 for optical and photocurrent images). Its optoelectronic and device characteristics are shown in Figure 2. First, we find that this device exhibits an EQE > 50% (Figure 2 (a)) with absorbance greater than 90% from approximately 500 nm to 600 nm. Spectral features such as the exciton resonances of $MoS_2$ and $WSe_2$ are well reproduced in the external quantum efficiency spectrum. In addition, we observe a maximum single-wavelength power conversion efficiency (PCE) of 3.4% under 740 W/cm$^2$ of 633 nm laser illumination (Figure 2 (b)). Since the high-performance device is electrically in parallel with other devices, typical macroscopically large spot size (~cm) AM 1.5G illumination measurements would yield device characteristics substantially different from the high-performing one. Thus, we estimated the AM 1.5G performance using extracted device parameters of a diode fit under laser illumination (see Supporting Information S2 for details). We estimate the AM 1.5G PCE of this device to be ~0.4%. This value is presently too low to be useful for photovoltaics, but the high EQE values reported here indicate promise for high efficiency devices, when device engineering efforts are able to also achieve correspondingly high open circuit voltages in van der Waals based photovoltaics.

Further measurements were performed at different laser powers under 633 nm laser illumination (Figure 3), yielding various power-dependent characteristics. Examination of the short-circuit current $I_{sc}$ yielded nearly linear dependence on laser power, as expected in ideal photovoltaic devices. The dashed blue line represents the fit to the expression $I_{sc} = AP^\tau$, where $A$ is a constant of proportionality, $P$ is the incident power, and $\tau$ represents the degree of nonlinearity in this device ($\tau = 1$ is the linear case).[20] We find that $\tau = 0.98$ in our device, indicating nearly linear behavior under short circuit conditions. In addition, in an ideal photovoltaic device, the open circuit voltage is expected to grow logarithmically with the input power, since $V_{oc} = \frac{nk_bT}{q}\ln\left(\frac{J_L}{J_{dark}} + 1\right) \approx \frac{nk_bT}{q}\ln\left(\frac{J_L}{J_{dark}}\right)$ for large illumination current densities $J_L$. Here, $J_{dark}$ is the dark current density, $n$ is the ideality factor, $k_b$ is the Boltzmann constant, $T$ is the temperature of the device, and $q$ is the fundamental unit of charge, so that $\frac{k_bT}{q} \approx 0.0258\ V$ at room temperature. In Figure 3 (b) we see that the experimental data match well with the diode fit (dashed black line, see Supporting Information S2 for fitting details), suggesting an ideality factor of $n = 1.75$ and a dark current density $J_{dark} = 0.65$ mA/(cm$^2$) assuming a $30\ \mu m \times 30\ \mu m$ device area. Also, since the power conversion efficiency (PCE) is given as $PCE = J_{sc}V_{oc}FF/P_{in}$, where $J_{sc}$ is the short circuit current density, $V_{oc}$ is the open circuit voltage, $FF$ is the fill fraction, and $P_{in}$ in the incident power density, we would expect the power conversion efficiency to scale roughly logarithmically as well. This is true for laser powers up to $\sim$740 W/cm$^2$ (Figure 3 (c)). However, for larger input power, the PCE decreases with increasing power. Such a drop in PCE can be attributed to series resistances in the device, either at the contacts or at the junction. This is corroborated by the match between the experimental data (dots) and the fitted expression (dashed line), yielding the diode fitting parameters in the lower right hand corner of the plot in Figure 3 (c). The fit for the $V_{oc}$ was simultaneously done with the PCE, therefore yielding the same set of parameters and a good match between experiment and extracted device parameters. Finally, we observed a decrease in the EQE at 633 nm with increasing power (Figure 3 (d)). Using the above fitted parameters, series resistance can only be used to partially explain a decrease in the EQE at higher powers. Thus, the additional decrease in EQE at higher powers may be due to the onset of carrier density-dependent nonradiative processes such as Auger or biexcitonic recombination which are not accounted for in the diode fit used above, where the dark current is fixed for all powers.

## Absorption in van der Waals heterostructures

We first investigate the absorption and optical properties of van der Waals heterostructures. We formed a heterostructure composed of hexagonal boron nitride (hBN)/ FLG/WSe$_2$/MoS$_2$/Au. The composite heterostructure has various regions (inset of Figure 4(a)), corresponding to different vertical heterostructures. Given the sensitivity of the performance of van der Waals materials to different environmental conditions and device fabrication procedures,[21] the samples fabricated here allow us to study optical and electronic features of different heterostructures in a systematic manner by probing specific heterostructures fabricated on the same monolithic substrate. This is enabled by the small spot size of our laser, which additionally allows us to properly normalize the spectral response without artificially including geometric factors (see Methods for details).

As an example, consider the optical response at the location of the blue dot in the inset of Figure 4 (a). The vertical heterostructure there is composed of 1.5 nm FLG/4 nm WSe$_2$/5 nm MoS$_2$/Au. This location can be probed spectrally for its absorption characteristics (Figure 4 (a)), revealing near-unity absorption in van der Waals heterostructures. The peaks at ~610, ~670, and ~770 nm correspond to the resonant excitation of the $MoS_2$ B exciton, $MoS_2$ A exciton, and $WSe_2$ A exciton, respectively.[22] On the other hand, the broad mode at ~550 nm corresponds to the photonic mode that leads to near-unity absorption.[23] Measurements of the absorption can be corroborated with electromagnetic simulations, unveiling both the accuracy between simulation and experimental results as well as the fraction of photon flux absorbed into individual layers of the heterostructure stack (Figure 4 (b)). Despite the near-unity absorption observed in the heterostructure stack, there is parasitic absorption in both the underlying gold substrate and in the few-layer graphene that accounts for ~20% of the total absorbance. Such parasitic absorption can be reduced by using a silver back reflector, as shown in Figure 4(c) and Figure 4 (d). We find that the simulated and measured absorbance is also in good agreement for the case of a silver back reflector. Thus, the optical response of a van der Waals heterostructure can be modelled accurately using full wave electromagnetic simulations and our method of measurement yields accurate and reliable results.

To note, the subwavelength dimension of the total heterostructure thickness is critical for achieving near-unity absorption. Indeed, the entire stack can be treated as a single effective medium, where small phase shifts are present between layers and therefore the material discontinuities are effectively imperceptible to the incident light (see Supporting Information S3 for details). Ultimately, the van der Waals heterostructure-on-metal behaves as a single absorbing material with effective medium optical properties. Therefore, as previously demonstrated, near-unity absorption at different wavelengths can be achieved for a semiconducting layer with the appropriate thickness[23,24] (~10-15 nm total thickness for TMD heterostructures).

## Carrier collection in van der Waals semiconductor junctions

As discussed above, another criterion for high EQE is efficient carrier collection. Given the large exciton binding energies in TMDs (~50-100 meV in the bulk),[25,26] the large internal electric field at the semiconductor heterojunction may play a role in exciton dissociation and subsequent carrier collection. Charge carrier separation in TMDs can be accomplished using either a pn junction or a Schottky junction, and we find that a pn heterojunction dramatically enhances the EQE when compared with a Schottky junction.

The heterostructure described in Figure 4 (a) and (b) can be probed as an optoelectronic device with the formation of a top electrode (see inset of Figure 5 (a)). Since the back reflector (gold) can simultaneously serve as a back contact to the entire vertical heterostructure, we can use this scheme to compare the electronic performance of various vertical heterostructures. Given the work function between $WSe_2$ (p-type) and Au, it is expected that a Schottky junction[27] will form between the two materials (See Figure 1 (c)), whereas $WSe_2$ (p-type) on top of $MoS_2$ (n-type) is expected to form a pn heterojunction.[4] High spatial resolution scanning photocurrent microscopy allows us to examine the two heterostructure devices in detail (Figure 5 (a)). We observe large photocurrent for the pn heterojunction geometry (yellow region) compared to the Schottky junction geometry (light blue region). The decrease of the photocurrent in the left-side of the yellow region in Figure 5(a) is due to shadowing from the electrical probes. A line cut of the spatial photocurrent map shown in Figure 5 (b) provides a clearer distinction between the two junctions, demonstrating ∼6x more photocurrent for the pn junction relative to the Schottky junction.

The photocurrent density is directly related to the external quantum efficiency and therefore the product of the absorbance and IQE. In order to quantitatively compare the *electronic* differences between the two junctions, we need to normalize out the different optical absorption in the two devices, *i.e.* compute the IQE of each device

$$IQE_{Exp}(\lambda) = \frac{EQE(\lambda)}{Abs(\lambda)} \qquad (1)$$

where $EQE(\lambda)$ and $Abs(\lambda)$ are the experimentally measured EQE and absorbance of their respective devices (Figure 5 (c) (i) and Figure 5 (d) (i)). A plot of the experimentally derived IQE (*i.e.* $IQE_{Exp}$) is shown in purple in Figure 5 (c) (ii) and Figure 5 (d) (ii). This plot also confirms that a pn junction geometry (with $IQE_{Exp}$∼40%) formed of van der Waals materials is more efficient for carrier collection than a Schottky junction geometry (with $IQE_{Exp}$ ∼10%).

Embedded in the above analysis is yet another convolution of the optical and electronic properties. As per Figure 4 (b), we found that absorption in FLG and Au accounted for ∼20% of the absorbance of the total heterostructure. Assuming very few photons absorbed in those layers ultimately are extracted as free carriers (*i.e.* $IQE_{Au} \approx IQE_{FLG} \approx 0$), the IQE defined above convolutes the parasitic optical loss with the electronic loss in the device.[28] Thus another useful metric we shall define is $IQE_{Active}$, the active layer IQE:

$$IQE_{Active}(\lambda) = \frac{EQE(\lambda)}{Abs(\lambda) - Abs_P(\lambda)} \qquad (2)$$

where the additional term $Abs_P(\lambda)$ corresponds to the parasitic absorption in the other layers of the device that do not contribute to current (*i.e.* Au and FLG in this device). Thus, $IQE_{Active}(\lambda)$ is a measure of the carrier generation and collection efficiency only in the *active* layer (*i.e.* WSe$_2$ and MoS$_2$) of the device and is *purely* an electronic efficiency as defined above. We shall use this quantity to accurately compare electronic geometries. Given the good agreement between simulations and experiment shown in Figure 4, a simple method of estimating the parasitic absorption described above is therefore through electromagnetic simulations. $IQE_{Active}$ of the Schottky and pn heterojunction geometries calculated with Eqn. 2 is shown in Figure 5 (c) (i) and Figure 5 (d) (ii)) with dotted green curves.

Analysis of these plots reveals several important points. First, $IQE_{Active}$ for the pn junction geometry is ∼3x higher than in the Schottky junction geometry when spectrally averaged. Though yet to be fully clarified, we attribute higher IQE in pn heterojunctions to the larger electric fields in a pn heterojunction that may lead to a higher exciton dissociation efficiency and consequently IQE. Second, compared to the IQE which included the parasitic absorption (purple dots in Figure 5 (c) (ii) and Figure 5 (d) (ii)), the active layer IQE curves (green dots) are spectrally flat within measurement error and calculations ($\delta IQE/IQE \approx 0.07$). Thus, the few broad peaks around the exciton energies of WSe$_2$ (~770 nm) and MoS$_2$ (~610 nm and ~670 nm) in $IQE_{Exp}$ are not attributed to, *e.g.*, resonant excitonic transport phenomena, but rather as a simple convolution of the optical and electronic effects when calculating the electronic IQE. In other words, consideration of parasitic absorption is critical when analyzing the electronic characteristics of thin optoelectronic devices. However, $IQE_{Exp}$ is still a useful metric, as it effectively sets a lower bound on the true IQE. Generally, we expect $IQE_{Exp} \leq IQE_{True} \leq IQE_{Active}$, as electromagnetic simulations tend to slightly overestimate the absorption when compared with experimental results. Thus in this paper, we shall plot both expressions when comparing different electronic device geometries. Finally, it is important to mention that an active layer IQE of ~70% is achieved in van der Waals heterostructures without complete optimization of the electronic configuration of the device, such as the band profiles and the specific choice of contacts. With careful electronic design, we suggest it may be possible to achieve active layer IQEs > 90%.

## Optically transparent contacts for carrier extraction

As another aspect of analysis, we studied the role of vertical carrier collection compared to lateral carrier collection in van der Waals heterostructures. Graphene and its few-layer counterpart can form a transparent conducting contact allowing for vertical carrier collection, in contrast to in-plane collection (see Figure 1 (d)). The strong, in-plane covalent bonds of van der Waals materials suggest that in-plane conduction may be favorable when contrasted with the weak out-of-plane van der Waals interaction. However, the length scale for carrier transport in-plane (~$\mu$m) is orders of magnitude larger than in the vertical direction (~nm). Therefore, transport in a regime in between these two limiting cases is not surprising.

Silver exhibits lower absorption in the visible than gold, suggesting it could be an optimal back reflector for photovoltaic devices, as seen in Figure 4. Thus, we contrast the case of in-plane and out-of-plane conduction concurrently with the presence of two different back reflectors that simultaneously function as an electronic back contact (gold vs. silver) to a pn heterojunction, as in Figure 6 (a) and (b). Optical and photocurrent images of the devices are shown in Supporting Information S1.

Our results in Figure 6 (c) and (d) show the distinctions between the various contacting schemes. In the case of both silver and gold, a transparent top contact such as few-layer graphene seems to enhance the carrier collection efficiency. This is particularly true in the case of silver, where $IQE_{Active}$ enhancements of ~5x is apparent. In the case of gold, the IQE is enhanced by about ~1.5x when parasitic absorption is taken into account. By analyzing the work functions of gold (~4.83 eV[29]) and silver (~4.26 eV[30]), along with the electron affinity of $MoS_2$ (~4.0 eV[31]), the Schottky-Mott rule suggests in both cases that a Schottky barrier should form equal to $\phi_B = \phi_M - \chi$,[32] where $\phi_B$ is the Schottky barrier height, $\phi_M$ is the work function of the metal, and $\chi$ is the electron affinity of the semiconductor. However, several reports[33–35] have indicated that gold appears to form an electrically Ohmic contact to $MoS_2$, which we observe here. Conversely, the above data suggests that silver and $MoS_2$ follow the traditional Schottky-Mott rule, leading to the formation of a small Schottky barrier of ~0.26 eV. Given that the energy barrier is about $10k_bT$, very few electrons can be extracted out of the pn heterojunction when silver is used as a back contact, leading to very low IQEs. By taking into account just the active layer (dashed lines), we see that gold is ~2x better as an electronic contact than silver.

Finally, we examine the role of vertical carrier collection on the I-V characteristics of the two devices (Figure 6 (e) and (f)). In the case of gold, we see purely an enhancement of the short circuit current with vertical carrier collection. On the other hand, vertical carrier collection for silver drastically increases both the short circuit current density and the open circuit voltage. This phenomenon is consistent with the previously described nature of gold (Ohmic) and silver (Schottky) contacts. Namely, on silver in the absence of a transparent top contact, due to both the Schottky barrier and the large in-plane propagation distance, carriers are collected with poor efficiency leading to a small $I_{sc}$. Consequently, a high recombination rate of the generated carriers which are inefficiently extracted leads to small $V_{oc}$ values. On the other hand, even in the absence of a top transparent electrode, gold enables efficient extraction of electrons from the pn heterojunction as an Ohmic contact. Thus, the short circuit current and open-circuit voltage in gold are higher compared to the silver back contact, even in the absence of a transparent electrode. When introducing few-layer graphene as a transparent top contact, the propagation distance is significantly reduced in the silver device and carriers can be extracted with much higher efficiency, leading to a large enhancement of both the current and voltage. Whereas for gold, the few-layer graphene enhances the already high carrier collection (yielding larger $I_{sc}$) but only has a negligibly small enhancement effect on the open-circuit voltage. Overall, these results demonstrate that vertical carrier collection plays a crucial role in high photovoltaic device performance in van der Waals heterostructures.

## Thickness dependence on charge collection efficiency

As a final point of analysis, we briefly examined the effect of thickness on $IQE_{Active}$ under vertical carrier collection. We compared the optoelectronic characteristics of a thicker pn heterojunction (11 nm hBN/1.5 nm FLG/4 nm $WSe_2$/9 nm $MoS_2$/Au) with a thinner pn heterojunction (1.5 nm FLG/4 nm $WSe_2$/5 nm $MoS_2$/Au). The experimentally measured absorbance and EQE are plotted in Figure S4 for reference. By normalizing out the differences in absorption between the pn junctions, we see a somewhat surprising result when we analyze the active layer IQE (dashed lines, Figure 7). In particular, despite the roughly 50% more length in active layer thicknesses (13 nm vs. 9 nm) and qualitatively different absorbance and EQE spectra, the thick pn junction exhibits nearly the same active layer IQE compared to the thin pn junction. In fact, it appears to be slightly more efficient, but this is within the error bar of the measurement and simulations ($\delta IQE/IQE \approx 0.07$, see Methods for details of errors). This observation is corroborated with the experimentally derived IQE (dotted curves, Figure 7), which has nearly the same spectrum between the two thicknesses, but differ in magnitude due to differences in parasitic absorption. This result suggests that in the ultrathin limit (~10 nm) of van der Waals heterostructures with vertical carrier collection, the IQE has a weak dependence on active layer thickness. This weak dependence may be due to a combination of increased scattering competing with charge transfer,[36,37] tunneling,[4,38,39] and exciton quenching[40,41] effects as the vdW heterostructure becomes thicker. The exact role of each of these effects, as well as possibly other effects, will require a new theoretical framework and experimental measurements to analyze their relative contributions to charge collection efficiency.

## CONCLUSIONS

Our results suggest important challenges that must be addressed to enable high photovoltaic efficiency. For example, despite the usefulness of gold as an electrical back contact, we found from electromagnetic simulations that it accounts for nearly 20% of the parasitic loss in the heterostructures reported here. Schemes using optically transparent carrier selective contacts could be used to avoid this parasitic optical loss. Another open question is the role and importance of exciton dissociation and transport. Indeed, the large exciton binding energies in transition metal dichalcogenides (~$50 - 100$ meV in the bulk)[25,26] suggests that a significant exciton population is generated immediately after illumination. However, it is not yet clear whether such an exciton population fundamentally limits the internal quantum efficiency of the device, posing an upper limit on the maximum achievable EQE in van der Waals materials based photovoltaic devices. Finally, the problem of open-circuit voltage must also be addressed. For example, the type-II band alignment between ultrathin $MoS_2$ and $WSe_2$ suggests a renormalized bandgap of ~$400 - 500$ meV,[42] given by the minimum conduction band energy and maximum valence band energy of the two materials. In accordance with the Shockley-Queisser limit, this would severely reduce the maximum power conversion efficiency attainable by a factor of ~3. Therefore, to achieve higher open circuit voltages, a monolithic device structure may be required to avoid low energy interlayer recombination states.

However, our results described here also suggest a different approach in addressing the optical and electronic considerations for ultrathin van der Waals heterostructures when compared with conventional photovoltaic structures. For example, our observation that ultrathin van der Waals heterostructures can be optically treated as a single effective medium is a regime of optics that is uncommon for the visible to near-infrared wavelengths analyzed in photovoltaic devices. Likewise, our observation of weak thickness dependence of the charge collection efficiency represents a realm of electronic transport that is also quite unconventional and unexplored when compared to traditional photovoltaic structures. Thus, the combination of the above observations may enable entirely different photovoltaic device physics and architectures moving forward.

To summarize, we have shown that external quantum efficiencies > 50% and active layer internal quantum efficiencies > 70% are possible in vertical van der Waals heterostructures. We experimentally demonstrated absorbance > 90% in van der Waals heterostructures with good agreement to electromagnetic simulations. We further used the active layer internal quantum efficiency to quantitatively compare the electronic charge collection efficiencies of different device geometries made with van der Waals materials. By further reducing parasitic optical losses and performing a careful study on exciton dissociation and charge transport while simultaneously engineering the band profiles and contacts, van der Waals photovoltaic devices may be able to achieve external quantum efficiencies > 90%. Our results presented here show a promising and exciting route to designing and achieving efficient ultrathin photovoltaics composed of van der Waals heterostructures.

## METHODS

### Metal Substrates Preparation

Atomically smooth metal substrates were prepared using the template stripping technique.[43,44] We prepared the substrates using polished silicon wafers (University Wafer) with native oxide and then cleaned the silicon substrates *via* sonication in acetone (10 minutes) followed by sonication in isopropyl alcohol (10 minutes). Samples were then blow dried with nitrogen gas before cleaning with oxygen plasma (5 minutes, 100 W, 300 mTorr under $O_2$ flow).

Metal was then deposited *via* electron beam evaporation on the polished and cleaned surface of the silicon wafer. For gold (Plasmaterials, 99.99% purity), base pressures of ~3e-7 Torr was achieved before depositing at 0.3 A/s. This continued until a thickness of ~20 nm was achieved. Then, the rate was slowly ramped to 1 A/s and then held there until a total thickness of 120 nm was reached. For silver (Plasmaterials, 99.99% purity), following McPeak *et al.*,[45] we deposited at a base pressure of ~3e-7 Torr at 40 A/s for a final thickness of 150 nm. After deposition of the metal, an adhesive handle was formed using a thermal epoxy (Epo-Tek 375, Epoxy Technology). 1 g of part A Epo-Tek 375 and 0.1 g of part B Epo-Tek 375 was mixed in a glass vial and was let to settle for ~30 min, Afterward, individual droplets of the mixture was added directly onto the metallic surface before placing cleaned silicon chips (~ 1 cm$^2$) on top. The droplet of epoxy was let to settle under the weight of the silicon chip before placing on a hot plate (~80 C) for 2 hours. Individual chips were then cleaved with a razor blade, forming the final substrate consisting of atomically smooth metal/thermal epoxy/silicon. Typical RMS surface roughness of the metal was < 0.3 nm using this technique (examined *via* AFM).

### Van der Waals Heterostructure Fabrication

The bottom-most layer of the van der Waals heterostructure (*e.g.* $MoS_2$) was directly exfoliated onto the metallic susbtrates prepared using the above technique. Exfoliation was performed using bulk crystals purchased from HQ Graphene using Scotch tape. Subsequent layers were formed using a visco-elastic dry transfer technique[46] using a home-built set-up at room temperature. Dry transfer was performed using PF-20-X4 Gel Film from Gel-Pak as the transparent polymer. Van der Waals materials were directly exfoliated onto the polymer using Nitto tape and then mechanically transferred onto the $MoS_2$/metal substrate. Samples were examined in the optical microscope during each layer of the process and an AFM scan was performed afterwards to extract out the thicknesses of individual layers. Thicknesses were then corroborated with optical measurements and calculations.

A top electrode was patterned using standard photolithography techniques. NR-9 1000 PY was used as a negative resist. The resist was spun at 5000 RPM for 55 s before baking at 150 C for 1 minute. A mask aligner with a pre-patterned mask was used to define the features and aligned on top of the van der Waals heterostructure. After exposure for ~18s under 10 mW of UV light ($\lambda = 365$ nm), the resist was post-baked at 105 C for 1 min and cooled to room temperature. Finally, the resist was developed using RD-6 developer for 10-15 seconds before rinsing in deionized water for 35 seconds. The sample was then blown dry with nitrogen and examined under an optical microscope.

Electron beam deposition was then used to form the top ring electrodes (10 nm Ti/90 nm Au). Base pressures of ~3e-7 Torr was achieved before the beginning of the deposition. For titanium, a deposition rate of 0.3 A/s was used for the entirety of 10 nm. Immediately afterwards, gold was deposited at a rate of 0.3 A/s for 15 nm. The rate was slowly ramped to 0.6 A/s for 10 nm, and then to 0.9 A/s for another 10 nm. At 35 nm of total gold thickness, the rate was finally ramped to 1.0 A/s until the total gold thickness was 90 nm. The resist was then lifted-off using heated acetone (40-45 C) for 30 minutes. If needed, the samples were sonicated in 5 s intervals in acetone to remove the resist. The sample was then rinsed in isopropyl alcohol and blow dried with nitrogen.

**Spatial Photocurrent Map and IV Measurements**

Samples were contacted on the top electrode and bottom metallic substrate using piezoelectric controlled probes (MiBots, Imina Technologies) with ~3 $\mu m^2$ tip diameter under a confocal microscope (Axio Imager 2 LSM 710, Zeiss) with a long working distance objective (50x, NA = 0.55). Samples were first checked for photoresponse using dark IV and white light illumination. Voltage sweeps were performed using a Keithley 236 source measure unit and in-house written scripts. High resolution spatial photocurrent maps were performed using the same confocal microscope with an automated stage. The microscope was modified to measure photocurrent maps. Diffraction limited laser light (~6 $\mu m^2$ spot size) was coupled in and focused to perform high resolution spatial photocurrent maps (< 1 $\mu m$ lateral resolution), and power-dependent IV measurements were performed at particular locations of the device using the spatial photocurrent maps. Illumination power was modified using neutral density filters in the microscope and the incident power was measured by a photodetector and cross-referenced with the EQE spectrum of the measured device.

**Spectral Absorbance and EQE Measurements**

Quantitative absorbance and external quantum efficiency measurements were performed using a home-built optical set-up. A supercontinuum laser (Fianium) was coupled to a monochromater to provide monochromatic incident light. A series of apertures and mirrors were used to collimate the beam before being focused on the sample with a long working distance (NA = 0.55) 50x objective to provide a small spot size (~1 $\mu m$ lateral resolution). Importantly, the small spot sizes allow us to probe individual regions on a particular sample. In addition, the relatively angle-insensitive light trapping structure[23] used in this work along with a small NA objective allowed us assume that the collected signal is close to the normal incidence response.

During all measurements, the light is first passed through a chopper (~103 Hz) and a small fraction was split into a photodetector connected to a lock-in amplifier. The other beam-split light is used for probing the sample which is eventually sent to a photodetector (for absorbance measurements) or the sample itself is used as a photodetector (for external quantum efficiency measurements). Thus, the sample or photodetector is connected to a second lock-in amplifier for homodyne lock-in detection.

For absorption measurements, the reflected signal was collected by the same 50x objective and passed through a beam splitter before being collected by a photodetector. The same spectral measurement was done with a calibrated silver mirror (Thorlabs) in order to obtain the absolute reflection spectrum. In the absence of transmission, the absorption is simply $Abs(\lambda) = 1 - R(\lambda)$. Reflection from the objective itself and other optical losses was subtracted as a background. As mentioned before, a reference spectrum was collected using a small amount of beam-split light at the same time as the sample, background, and mirror scans to account for power fluctuations in the laser beam between scans. As a second reference, the metallic back substrate was measured during all absorption scans to check if the normalization was accurate.

For external quantum efficiency measurements, the sample itself was used as a photodetector. The top ring electrode and bottom metallic substrate was probed using MiBots. Laser light was then focused on a particular spot and the current was collected by the probes and sent through a lock-in amplifier for homodyne detection, as in the reflection spectrum case. After measurement of the current signals from the sample, another spectral scan was performed with the optical system in the same configuration using a NIST calibrated photodetector (818-ST2-UV/DB, Newport). Power fluctuations between scans were again accounted for by using a small amount of beam-split light and sending it to a photodetector. The measured currents were normalized to this photodetector's current before being normalized to the calibrated photoresponse to yield the absolute EQE.

Despite the various steps of calibration used for normalization, we still estimate measurement errors of $\delta Abs/Abs \approx 0.02$ and $\delta EQE/EQE \approx 0.05$ stemming from the assumption of normal incidence for both absorption and external quantum efficiency measurements while using a NA = 0.55 objective, fluctuations in the laser power during the measurement, and sample contact stability. In addition, we have observed in our laser that there is relatively little power for $\lambda < 450$ nm. Additionally, there is relatively high absorbance in the 50x objective for $\lambda > 700$ nm. Combined with the fact that the simulated parasitic absorption accounts for a larger fraction of the total absorption for $\lambda < 450$ nm and $\lambda > 700$ nm, the significantly noisier spectra in the active layer IQE at these wavelengths can be attributed to the factors described above.

**Electromagnetic Simulations**

Calculations were performed using the transfer matrix method[20] with optical constants taken from literature for each of the transition metal dichalcogenides (TMDs).[22] We assumed that for the TMD thicknesses analyzed in this paper, their optical response can be represented by the bulk optical permittivities. Permittivities of Ag and Au were taken from McPeak[45] and Olman,[47] respectively. The optical response of few-layer graphene was assumed to be like graphite, with its dielectric constant taken from Djurisic.[48] Hexagonal Boron Nitride (hBN) was assumed to be a lossless, non-dispersive dielectric in the visible with refractive index of $n = 2.2$.[49]

Given that there is sample-to-sample variation of the dielectric constant, it is likely that the literature values of the dielectric constant differ from the samples measured here. This difference we estimate leads to absorption simulation errors of ~5%. Assuming this is true, the estimated error for the active layer IQE can be approximated as:

$$\frac{\delta IQE}{IQE} \approx \sqrt{\left(\frac{\delta Abs}{Abs}\right)^2 + \left(\frac{\delta Abs_P}{Abs_P}\right)^2 + \left(\frac{\delta EQE}{EQE}\right)^2}$$

which is about 7%.

## ASSOCIATED CONTENT:

**Supporting Information:**

Additional experimental data, calculations and analysis accompany this paper. This material is available free of charge *via* the Internet at http://pubs.acs.org

## AUTHOR INFORMATION

**Corresponding Author:**

*Harry A. Atwater, E-mail: haa@caltech.edu

**ORCID:**

Joeson Wong: 0000-0002-6304-7602

Michelle C. Sherrott: 0000-0002-7503-9714

**Author Contributions:**

J.W. and D.J. prepared the samples and fabricated the devices. J.W. and A. R. D. performed the calculations. J.W., D.J., and G.T. performed the measurements. K.T. and M.C.S. assisted with sample preparation and fabrication. H.A.A. supervised over all the experiments, calculations, and data collection. All authors contributed to data interpretation, presentation, and preparation of the manuscript.

**Notes:**

The authors declare no competing financial interests.

## ACKNOWLEDGEMENTS:

This work is part of the "Light-Material Interactions in Energy Conversion" Energy Frontier Research Center funded by the U.S. Department of Energy, Office of Science, Office of Basic Energy Sciences under Award Number DE-SC0001293. D.J., A.R.D., and M.C.S. acknowledge additional support from the Space Solar Power project and the Resnick Sustainability Institute Graduate and Postdoctoral Fellowships. A.R.D. also acknowledges support in part from the Kavli Nanoscience Institute Postdoctoral Fellowship. G.T. acknowledges support in part from the Swiss National Science Foundation, Early Postdoc Mobility Fellowship n. P2EZP2_159101. J.W. acknowledges support from the National Science Foundation Graduate Research Fellowship under Grant No. 1144469. K.T. would like to thank the Caltech SURF program and the Northrop Grumman Corporation for financial support.

## FIGURES:

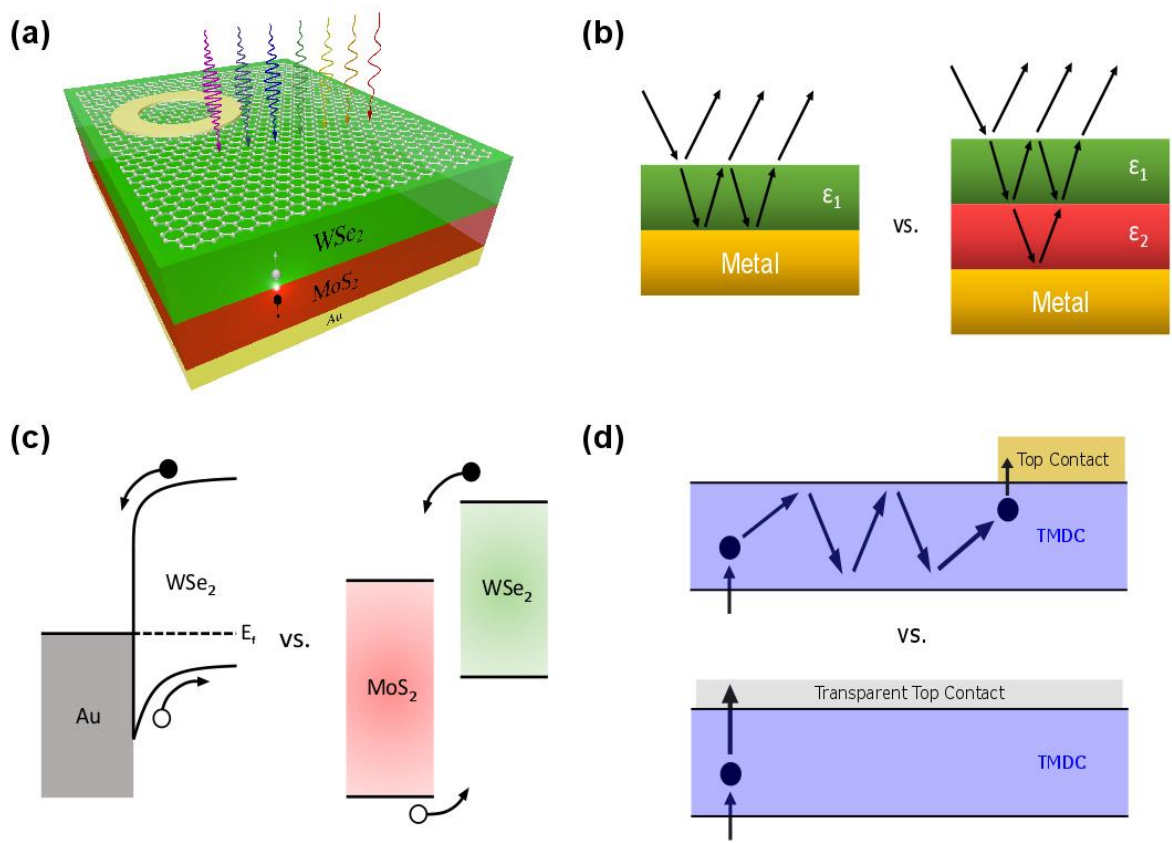

**Figure 1.** Achieving High External Quantum Efficiency in van der Waals heterostructures: **(a)** A schematic of the van der Waals device stack where nanophotonic light trapping combined with efficient exciton dissociation and carrier collection yields EQEs >50%. **(b)** A schematic of comparing near-unity absorption in a single semiconducting layer on metal with a heterostructure of different semiconductors on metal. **(c)** A schematic of comparing a pn heterojunction with a Schottky junction for exciton dissociation and charge carrier separation in van der Waals materials. **(d)** A schematic of comparing vertical and lateral carrier collection schemes in van der Waals materials.

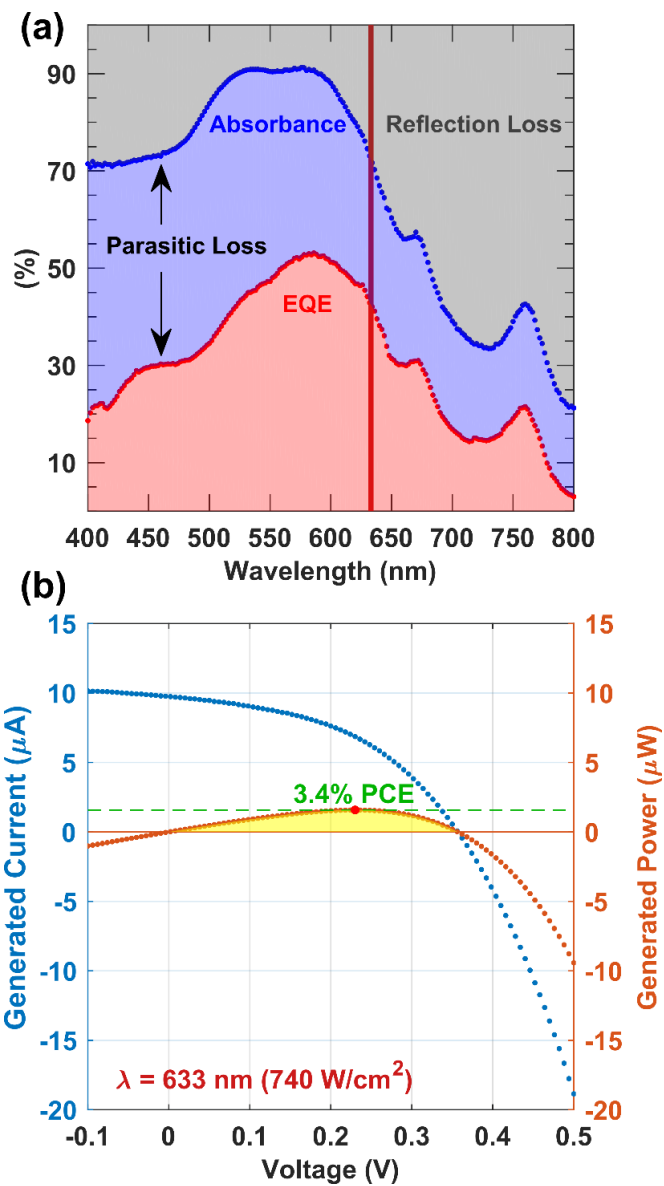

**Figure 2**. Optoelectronic performance characteristics: **(a)** Spectral characteristics of the experimentally measured absorbance (blue) and external quantum efficiency (red). The vertical solid line indicates the excitation wavelength (633 nm) for the measurements in (b) and Figure 3. The grey region indicates loss in photocurrent from the reflected photons. **(b)** I-V (light blue) and power-voltage (orange) characteristics of the device, excited at $\lambda = 633$ nm with ~45 $\mu$W incident power with a spot size area of ~6 $\mu$m$^2$. We observe a maximum single wavelength power conversion efficiency of 3.4%. The yellow region indicates generated power from the device.

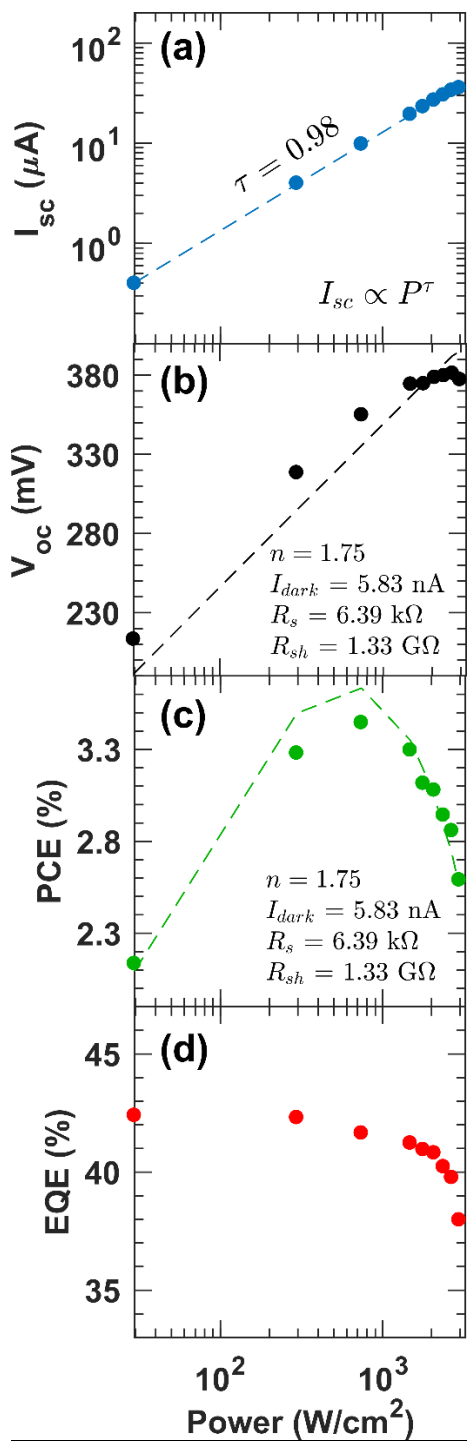

**Figure 3.** <u>Power dependent device characteristics</u>: Power dependent device characteristics at $\lambda = 633$ nm excitation for the **(a)** short circuit current (light blue), **(b)** open circuit voltage (black), **(c)** maximum power conversion efficiency (green), and **(d)** external quantum efficiency (red). The dashed lines in (a), (b), and (c) correspond to fits. The area of the spot size of the laser in all of the above measurements is estimated to be $\sim 6$ $\mu m^2$.

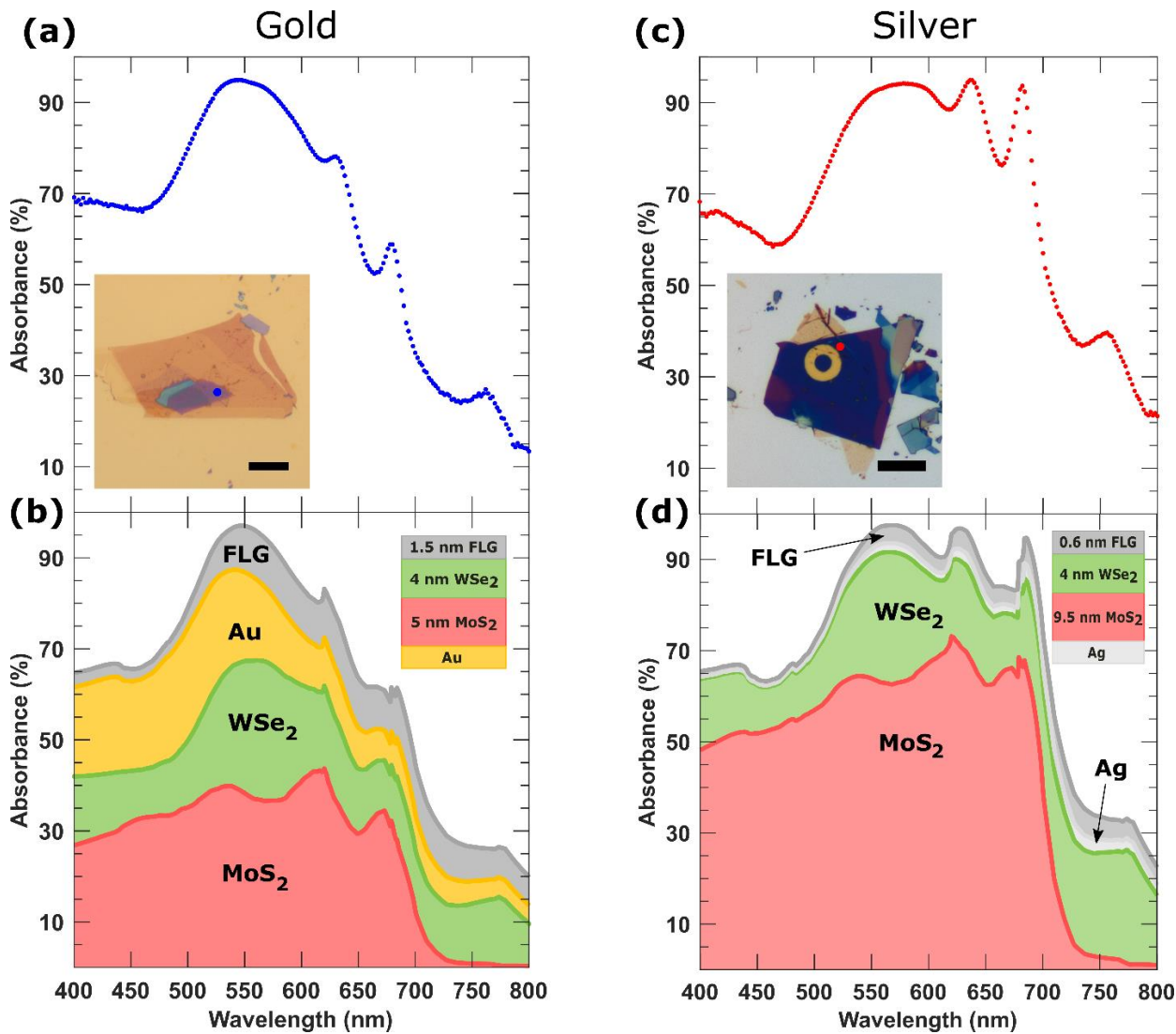

**Figure 4.** Absorbance in van der Waals heterostructures: **(a)** Experimentally measured absorbance of the 1.5 nm FLG/4 nm WSe$_2$/5 nm MoS$_2$/Au stack as a function of wavelength. The inset is an optical micrograph of the fabricated van der Waals heterostructure (scale bar = 20 $\mu$m) with the blue dot corresponding to the spot of spectral measurement. **(b)** The simulated absorbance of the structure in (a) partitioned into the fraction of absorbance going into individual layers of the heterostructure stack. The inset is a cross-sectional schematic of the simulated and measured heterostructure. **(c)** Experimentally measured absorbance of the 0.6 nm FLG/4 nm WSe$_2$/9.5 nm MoS$_2$/Ag stack as a function of wavelength. The inset is an optical micrograph of the fabricated van der Waals heterostructure (scale bar = 20 $\mu$m) with the red dot corresponding to the spot of spectral measurement. **(d)** Same as in (b) except the simulated absorbance is for the sample fabricated on silver as shown in (c).

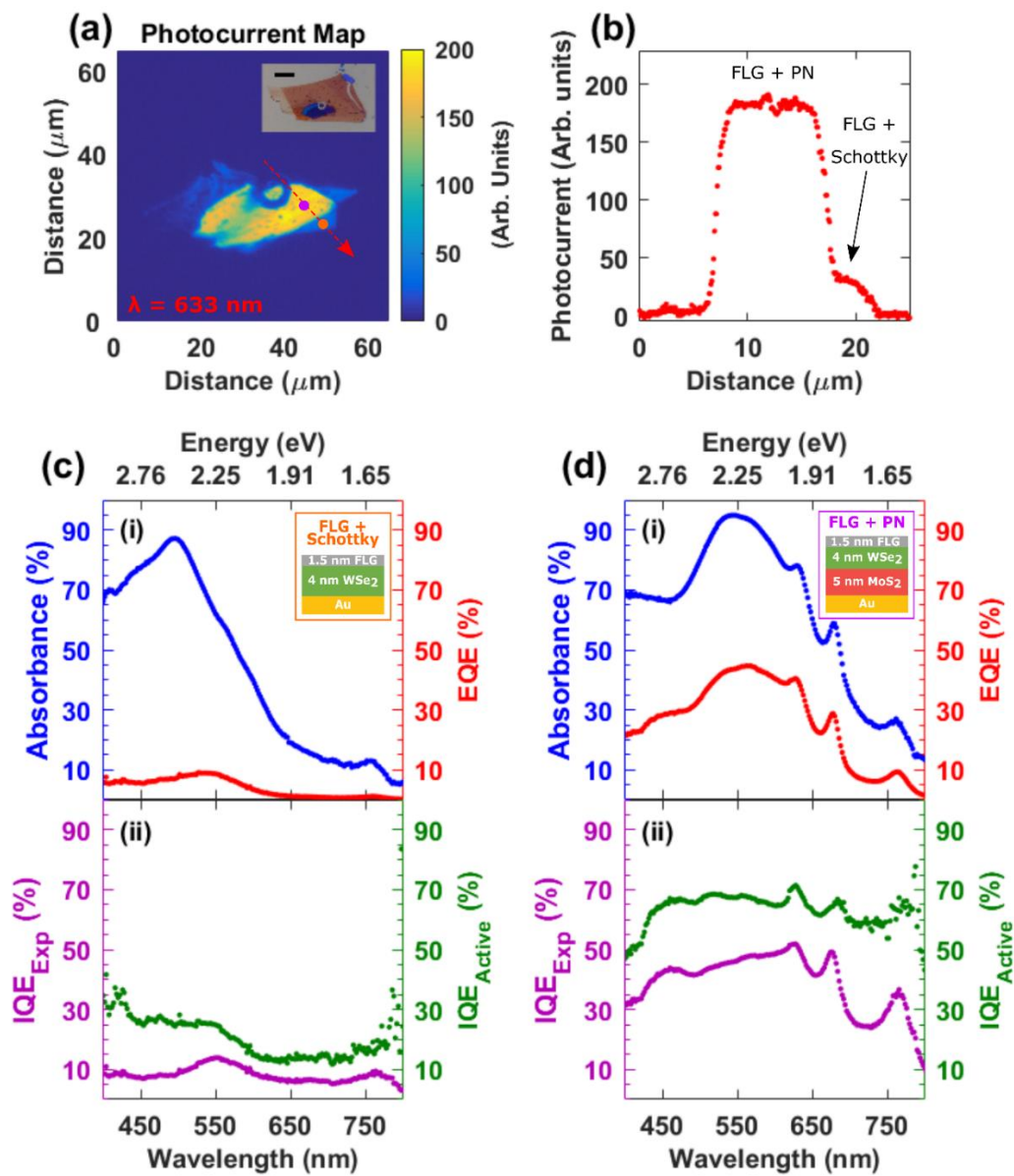

**Figure 5.** Charge transport and collection in vertical PN and Schottky junction geometries: **(a)** Spatial photocurrent map of the fabricated van der Waals heterostructure device using a 633 nm laser excitation. The inset is an optical image of the device (scale bar = 20 $\mu$m). **(b)** The line profile of the dotted red line arrow in (a), illustrating the different photocurrent intensities depending on the device geometry (Schottky and pn junction). **(c) (i)** Experimentally measured spectral characteristics of the absorbance (blue) and external quantum efficiency (red) in the 1.5 nm FLG/4 nm WSe$_2$/Au (Schottky geometry) device along with the **(ii)** experimentally derived internal quantum efficiency (purple) and the calculated active layer internal quantum efficiency (green). The inset is a cross-sectional schematic of the measured device, at the orange dot in (a). **(d)** Same as in (c) except with a 1.5 nm FLG/4 nm WSe$_2$/5 nm MoS$_2$/Au (pn geometry) device. The inset is a cross-sectional schematic of the measured device, at the purple dot in (a).

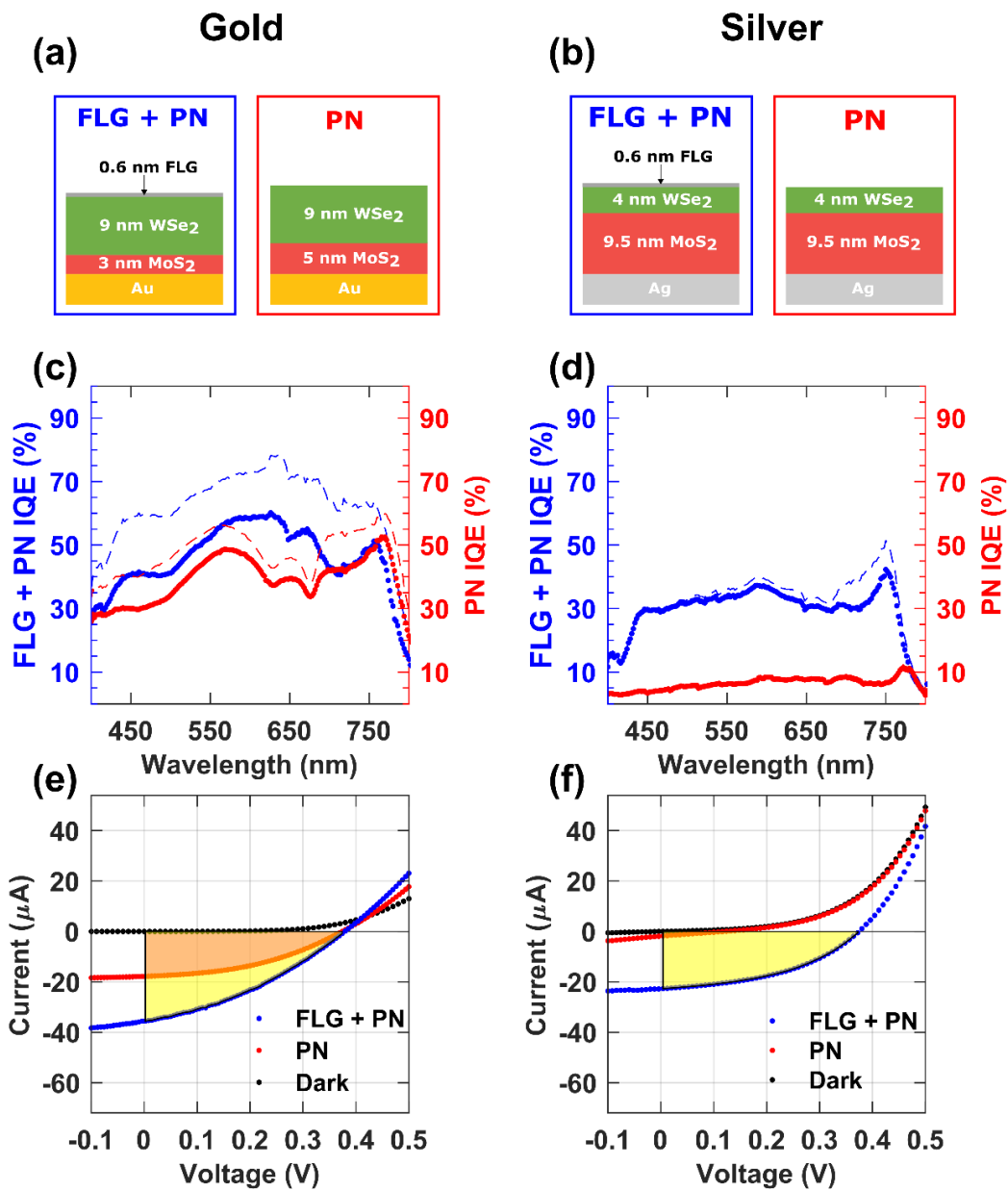

**Figure 6.** Few-layer graphene as a transparent top contact: **(a)** Cross-sectional schematic of the two structures (with and without few layer graphene) compared on gold and **(b)** silver back reflectors. **(c)** Experimentally derived internal quantum efficiency (dots) and active layer internal quantum efficiency (dashed line) for a pn junction geometry with (blue) and without (red) few layer graphene on gold. **(d)** is the same as (c) except on silver, corresponding to the sample shown in (b). **(e)** I-V curves of a pn junction geometry with (blue) and without (red) few-layer graphene under 633 nm (~180 $\mu$W) laser illumination on a gold and **(f)** silver substrate. The shaded yellow and orange regions correspond to where there is a net generated power in the device.

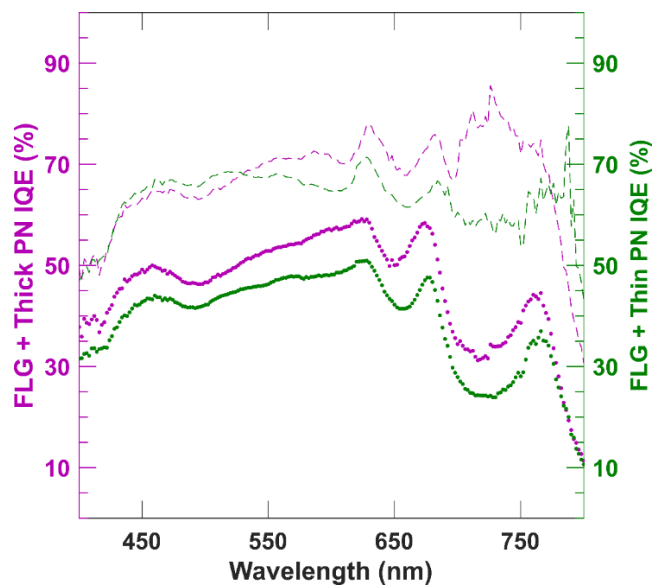

**Figure 7.** Thickness dependence on charge collection efficiency: The estimated active layer (dashed lines) and experimentally derived (solid dots) internal quantum efficiency of the thin pn junction device (1.5 nm FLG/4 nm WSe$_2$/5 nm MoS$_2$/Au, green) and the thick pn junction device (11 nm hBN/1.5 nm FLG/4 nm WSe$_2$/9 nm MoS$_2$/Au, purple).

Supporting Information for

# High Photovoltaic Quantum Efficiency in Ultrathin van der Waals Heterostructures

*Joeson Wong[1‡], Deep Jariwala[1,2‡], Giulia Tagliabue[1,4], Kevin Tat[1], Artur R. Davoyan[1,2,3], Michelle C. Sherrott[1,2] and Harry A. Atwater[1,2,3,4*]*

[1]Department of Applied Physics and Materials Science, California Institute of Technology, Pasadena, CA-91125, USA

[2]Resnick Sustainability Institute, California Institute of Technology, Pasadena, CA-91125, USA

[3]Kavli Nanoscience Institute, California Institute of Technology, Pasadena, CA-91125, USA

[4]Joint Center for Artificial Photosynthesis, California Institute of Technology, Pasadena, CA-91125, USA

\*  Corresponding author: Harry A Atwater (haa@caltech.edu)
‡ These authors contributed equally

## S1. Optical and Scanning Photocurrent Images of Samples

Optical and scanning photocurrent measurements were taken on all samples analyzed in this work, presented in Figure S1. Sample 1 is the high-performance device for which results are presented in Figure 2 and 3 of the main text. Results presented in Figure 5 and 7 correspond to Sample 2. Figure 6 draws comparisons between Samples 1 and 3, and Figure 4 draws comparisons between Samples 2 and 3.

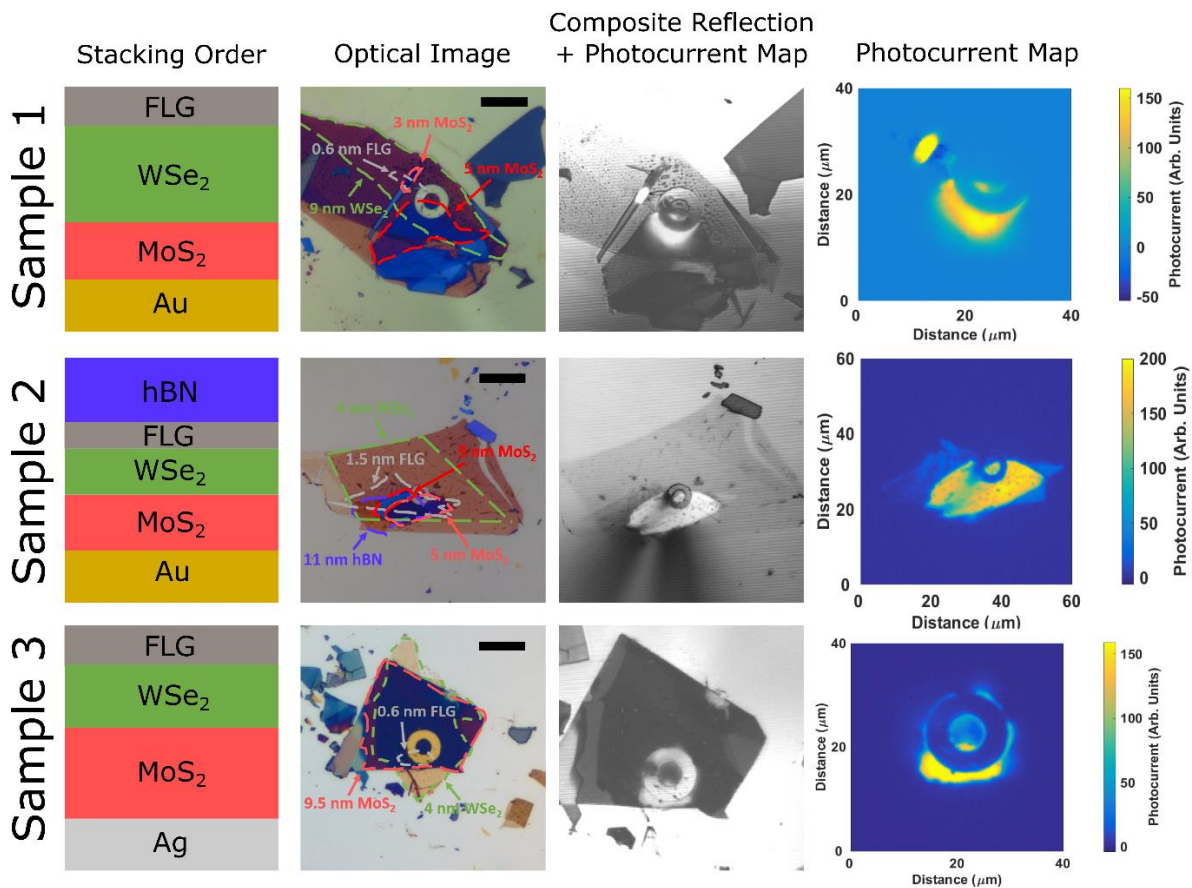

**Figure S1.** Sample Images: The heterostructure designs, optical images, composite reflection & photocurrent maps, and the photocurrent maps for all the samples analyzed in this paper. The outlines in the optical images correspond to specific materials with the appropriate thickness and materials labeled (scale bar = 20 $\mu m$). The composite reflection and photocurrent map is made by superimposing a reflection mode scan with a photocurrent scan ($\lambda = 633$ nm). The bright white regions in the composite image correspond to high photocurrent.

## S2. Diode equation fitting and simulated AM 1.5G

A diode model is commonly used to examine the characteristics of photovoltaic devices. Here, we assume a single diode model with a series and shunt resistance as a simple model to understand the photovoltaic device characteristics:

$$I(V) = I_{dark}\left(\exp\left(\frac{q(V - IR_s)}{nk_bT}\right) - 1\right) + \frac{V - IR_s}{R_{sh}} - I_L \qquad (1)$$

where $I_{dark}$ is the dark current, $q$ is the fundamental charge constant ($1.602 \times 10^{-19}$ C), $n$ is the ideality factor, $k_b$ is the Boltzmann constant ($1.38 \times 10^{-23}$ J K$^{-1}$), $T$ is the thermodynamic temperature (300 K, for this case), $R_s$ is the series resistance, $R_{sh}$ is the shunt resistance, and $I_L$ is the generated current from the photovoltaic effect under illumination. Here, $V$ is the applied voltage and $I$ is the measured current. At short circuit, $V = 0$ and $I \to I_{sc}$. Thus,

$$I_L = I_{dark}\left(\exp\left(\frac{-qI_{sc}R_s}{nk_bT}\right) - 1\right) - \frac{I_{sc}R_s}{R_{sh}} - I_{sc} \qquad (2)$$

For the case $R_s = 0$, we recover the usual expression $I_L = -I_{sc}$. We use the above two expressions along with the measured short circuit current $I_{sc}$ to perform a four parameter $(n, I_{dark}, R_s, R_{sh})$ fit to the open circuit voltage $V_{oc}$ and the power conversion efficiency $\eta = P_{device}/P_{input}$ as a function of input power. Here, we have explicitly measured the input power of the laser illumination. The fitted parameters are listed in Figure 3 (b) and Figure 3 (c) in the main manuscript, and are used to generate the dashed lines in those plots. Note that we use the same fitted parameters for both data sets. It is also important to note that by fitting the parameters under illumination at various powers, we expect the fitted parameters to represent primarily the device characteristics that are probed by laser illumination, and not all the other devices that are in parallel (which would be the case if we fitted to the dark IV).

To estimate the power conversion efficiency under AM 1.5G illumination for the particular device, we use the expression:

$$I_{sc} = -qA \int_{400\,nm}^{800\,nm} EQE_{exp}(\lambda) S_{AM\,1.5G}(\lambda) d\lambda \qquad (3)$$

where $q$ is the fundamental charge constant ($1.602 \times 10^{-19}$ C), $A$ is the estimated active area, $EQE_{exp}$ is the experimentally measured EQE for the device, and $S_{AM\,1.5G}$ is the solar photon flux (photons m$^{-2}$ s$^{-1}$ nm$^{-1}$). Using the above fitted parameters and the calculated $I_{sc}$, we can simulate the $I(V)$ characteristics of the device. We take $J(V) = I(V)/A$ and calculate the power conversion efficiency ($\eta$) as

$$\eta = \frac{J_m V_m}{\int_0^\infty \left(\frac{hc}{\lambda}\right) S_{AM\,1.5G}(\lambda) d\lambda} \qquad (4)$$

Where $J_m$, $V_m$ is the current density and voltage at the maximum power point, respectively, and the denominator of the above expression represents the total incident power of solar irradiation ($S = \int_0^\infty \left(\frac{hc}{\lambda}\right) S_{AM\,1.5G}(\lambda) d\lambda = 1000$ W m$^{-2}$). We plot this as a function of estimated active area $A$ in Figure S2 (a). Note that with increasing estimated active area, we observe an increase in the power conversion efficiency.

Here, the active area effectively reduces the dark current density $J_{dark} = I_{dark}/A$ for increasing $A$, and therefore leads to a concentration-like effect on the power conversion efficiency. Thus, there is a logarithmic dependence of $\eta$ on the active area $A$ and therefore $\eta$ varies weakly with $A$. Moreover, the above analysis for $A$ also allows us to estimate the appropriate area for the simulated device performance, as this is not the area under illumination, but rather the area from which dark current, series resistance, and shunt resistance contribute to the total measured current (*i.e.* the total sample size). We estimate this area to be in the range of $20^2 - 40^2 \ \mu m^2$ from the optical image (Figure S1) and plot the J-V characteristics assuming a $30 \times 30 \ \mu m^2$ active area below (Figure S2 (b)). Typical photovoltaic figures of merit are also shown. We achieve $J_{sc} > 8$ mA/cm$^2$ under 1 sun illumination. This value depends only on the experimentally measured EQE and does not depend on any fitting parameters, as evident in Eqn. 3. However, the expected $V_{oc}$ and $FF$ are sub-optimal, due to the type-II band alignment and high series resistance of the device. Thus, despite having fairly large short circuit current densities, device performance is limited primarily by the open circuit voltage and fill fraction, leading to an overall predicted $\eta_{AM\,1.5G} \approx 0.4\%$.

The above analysis differs from the typical experimental scenario where we estimate the input power as $P_{input} = SA$, where $S = 1000$ W m$^{-2}$ and $A$ is the illumination area. Thus, the experimental efficiency is given as $\eta = \frac{P_{m,exp}}{P_{input}}$, where $P_{m,exp}$ is the maximum power of the experimentally measured device. In the experimental case, $A$ is optimally the solar illumination area through some well-calibrated aperture.[1] In this case, the power conversion efficiency is inversely proportional to the estimated active area and therefore leads to larger $J_{sc}$ and $\eta$ for smaller $A$. This is a common source of error in estimating $\eta$ for small devices, as $|\delta\eta|/\eta = |\delta A|/A$, with the error in efficiency $\delta\eta$ depending linearly with the error in active area estimation $\delta A$. Particularly for micron and nano-scale devices such as in van der Waals materials, particular care must be taken to avoid errors in measuring and calculating the power conversion efficiency, as discussed by Snaith *et al.* in ref. 1. Here, we show a distribution of efficiencies based on our active area estimation, leading to AM 1.5G power conversion efficiencies between 0.25% to 0.5%. For our above calculation methodology, we can derive the error dependence to be roughly $\frac{|\delta\eta|}{\eta} \approx \frac{|\delta A|}{A}\left(\frac{nk_bT}{qV_{oc}}\right)$, where the extra factor of $nk_bT/(qV_{oc})$ comes from the dependence of $\eta$ with an estimated $V_{oc}$, rather than $J_{sc}$. The low values of absolute efficiency and logarithmic dependence on active area using our calculation methodology imply a weak dependence of the error on estimated active area, and thus suggests our calculated performance is a reasonable estimate for an experimental AM 1.5G measurement.

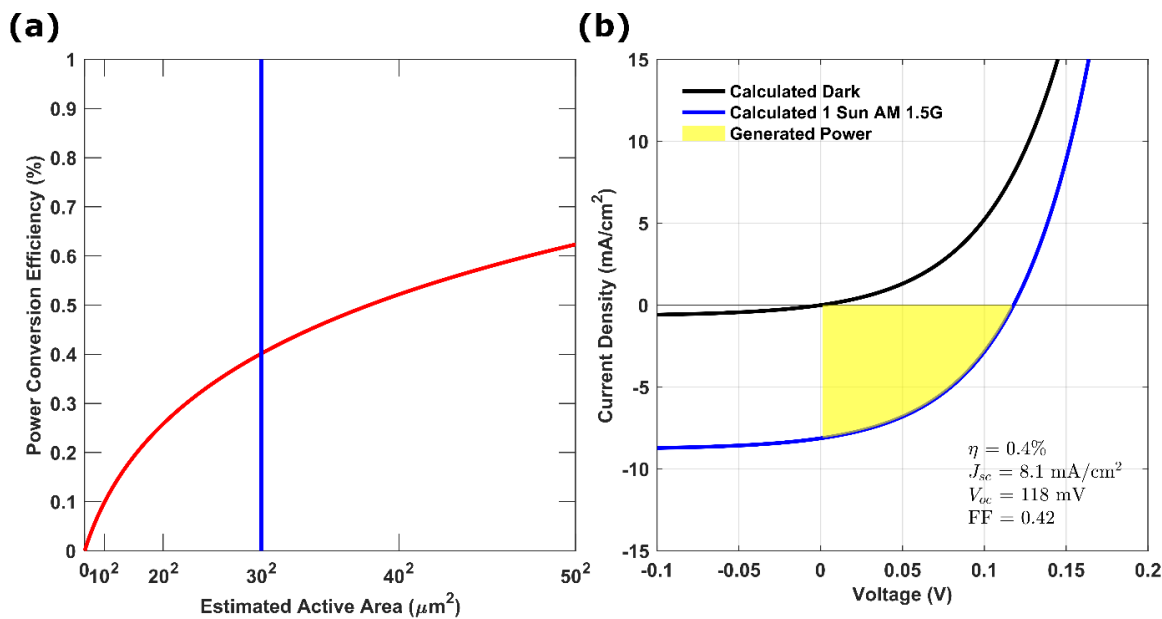

**Figure S2**. <u>Estimated 1 Sun AM 1.5G performance</u>: **(a)** Estimated 1 Sun AM 1.5G power conversion efficiency of the device measured in Figure 2 and 3 of the main manuscript as a function of estimated active area. The blue line corresponds to a $30 \times 30$ $\mu m^2$ estimated active area used for the plot in (b). **(b)** The estimated J-V curve of the device studied in Figure 2 and 3 of the main manuscript in the dark (black line) and under 1 Sun AM 1.5G illumination (blue) assuming a $30 \times 30$ $\mu m^2$ active area. Estimated device characteristics are in the bottom right-hand corner of the plot.

## S3. Effective medium theory in ultrathin heterostructures

Given that the samples considered in this work are deep sub-wavelength ($\sim$10-15 nm $\ll \sim$ 500 nm), it is expected that an effective medium theory can be employed. An effective medium dielectric constant was calculated using

$$\varepsilon_{EMT}(\lambda) = \frac{\sum_{j=1}^{N} \varepsilon_j(\lambda) t_j}{\sum_{j=1}^{N} t_j} \tag{5}$$

where $\varepsilon_{EMT}$ is the effective medium dielectric constant, $\varepsilon_j$ is the dielectric constant of the $j$th layer, $t_j$ is the thickness of the $j$th layer, and $N$ is the total number of layers, excluding the metal substrate. Figure S3 (a) shows the effective dielectric constant calculated for the sample studied in Figure 4 (a) and (b) of the main paper. Optically, the dielectric constant can be considered as a single absorbing dielectric layer. The validity of this approach is shown by calculating the absorption spectrum using the effective medium dielectric constant and all the individual dielectric components in a stack (Figure S3 (b)). Thus, one would expect similar non-trivial phase shifts at the absorbing dielectric – metal interface,[2] even when several layers are incorporated. In the deep sub-wavelength regime, the phase shifts between different van der Waals heterostructures are unimportant as the light will average over these effects. Therefore, near-unity absorption is also achievable in ultrathin van der Waals heterostructures, as has been demonstrated for single-component absorbers on a reflective surface.[3]

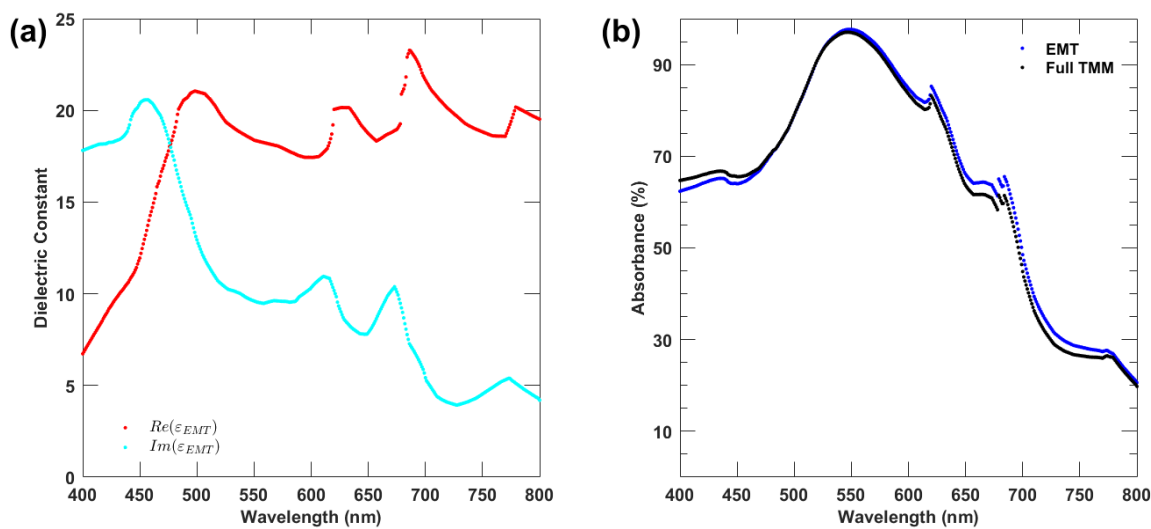

**Figure S3.** <u>Effective medium theory in ultrathin van der Waals heterostructures</u>: **(a)** The real (red) and imaginary (cyan) part of the effective medium theory (EMT) dielectric constant of the gold sample studied in Figure 4 of the main paper (1.5 nm FLG/4 nm WSe$_2$/5 nm MoS$_2$/Au). **(b)** The absorption spectrum calculated using the effective medium dielectric constant (blue dots, 10.5 nm EMT material/Au) versus the absorption spectrum calculated using all the dielectric components (black dots, 1.5 nm FLG/4 nm WSe$_2$/5 nm MoS$_2$/Au).

## S4. Absorbance and EQE Plots of Thick and Thin PN Junctions

Despite the vastly different absorbance and EQE spectra between the thick and thin pn heterojunction devices (Figure S4), along with the differences in active-layer thicknesses (13 nm and 9 nm, respectively), the active-layer and experimental IQE response of the two devices are comparable (Figure 7 of the main manuscript). This suggests competing effects in the generation and collection of carriers[4–10] in the ultrathin limit. Further experimental studies on carrier transport corroborated with theoretical models are needed to understand the role this may have on ultrathin photovoltaic devices made of van der Waals materials.

**(a)** **(b)**

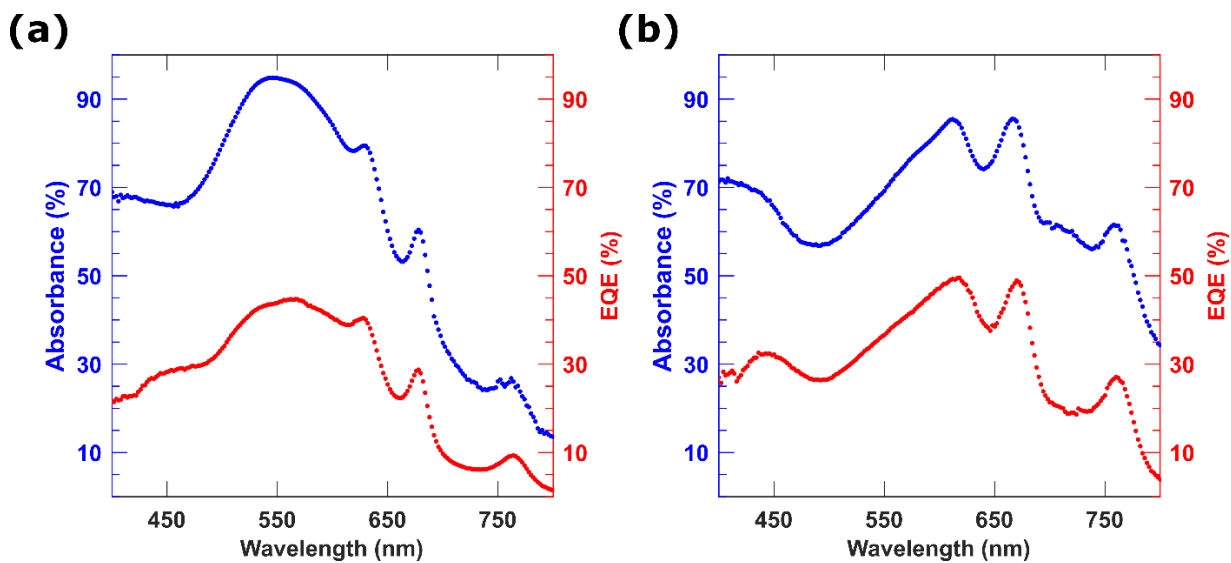

**Figure S4**. <u>Absorbance and EQE of Thick and Thin PN Junctions:</u> **(a)** Experimentally measured absorbance (blue) and EQE (red) of the thin pn heterojunction (1.5 nm FLG/4 nm WSe$_2$/5 nm MoS$_2$/Au). **(b)** Same as in (a) except for a thick pn heterojunction (11 nm hBN/1.5 nm FLG/4 nm WSe$_2$/9 nm MoS$_2$/Au).